\documentclass[acmsmall,screen]{acmart}

\AtBeginDocument{%
  }

\setcopyright{acmlicensed}
\copyrightyear{2025}
\acmYear{2025}
\acmDOI{XXXXXXX.XXXXXXX}

\acmJournal{TOSEM}
\acmVolume{0}
\acmNumber{0}
\acmArticle{0}
\acmMonth{8}

\usepackage{array}
\usepackage[noend]{algpseudocode}
\usepackage{algorithmicx,algorithm}
\usepackage{multirow}
\usepackage{tcolorbox}
\usepackage{float}
\usepackage{array}

\begin{document}

\title{Knowledge-Guided Prompt Learning for Request Quality Assurance in Public Code Review}

\author{Lin Li}
 \affiliation{
   \institution{School of Computer Science and Artificial Intelligence, Wuhan University of Technology}
   \city{Wuhan}
   \country{China}
 }
\email{cathylilin@whut.edu.cn}

\author{Xinchun Yu}
\authornote{Corresponding author}
\affiliation{
  \institution{School of Computer Science and Artificial Intelligence, Wuhan University of Technology}
  \city{Wuhan}
  \country{China}}
\email{evenyxc@whut.edu.cn}

\author{Xinyu Chen}
\affiliation{
  \institution{China Mobile Design Institute Co., Ltd.}
  \city{Wuhan}
  \country{China}}
\email{chenxinyu@cmdi.chinamobile.com}

\author{Peng Liang}
\affiliation{
  \institution{School of Computer Science, Wuhan University}
  \city{Wuhan}
  \country{China}}
\email{liangp@whu.edu.cn}

\renewcommand{\shortauthors}{}

\begin{abstract}
Public Code Review (PCR) is developed in the Software Question Answering (SQA) community, assisting developers in exploring high-quality and efficient review services. 
Current methods on PCR mainly focus on the reviewer's perspective, including finding a capable reviewer, predicting comment quality, and recommending/generating review comments. 
However, it is not well studied that how to satisfy the review necessity requests posted by developers which can increase their visibility, which in turn acts as a prerequisite for better review responses. 
To this end, we propose \underline{K}nowledge-guided  \underline{P}rompt learning for  \underline{P}ublic  \underline{C}ode  \underline{R}eview (KP-PCR) to achieve developer-based code review request quality assurance (i.e., predicting request necessity and recommending tags subtask). Specifically, we reformulate the two subtasks via 1) text prompt tuning which converts both of them into a Masked Language Model (MLM) by constructing prompt templates using hard prompt; and 2) knowledge and code prefix tuning which introduces knowledge guidance from fine-tuned large language models by soft prompt, and uses program dependence graph to characterize code snippets. Finally, both of the request necessity prediction and tag recommendation subtasks output predicted results through an answer engineering module. 
In addition, we further analysis the time complexity of our KP-PCR that has lightweight prefix based the operation of introducing knowledge guidance.
Experimental results on the PCR dataset for the period 2011-2023 demonstrate that our KP-PCR outperforms baselines by 2.3\%-8.4\% in the request necessity prediction and by 1.4\%-6.9\% in the tag recommendation. The code implementation is released at \url{https://github.com/WUT-IDEA/KP-PCR}. 
\end{abstract}



\maketitle

\section{Introduction}\label{introduction}
Code review can be conducted either within a team or by public developers. As an application form of traditional code review~\cite{shan2022using,chen2022code,li2023code,rahman2017predicting,pandya2022corms}, code review within a team, serves as an essential quality assurance mechanism in mature and successful commercial and Open Source Software (OSS) projects~\cite{bosu2013impact}. However, due to its labor-intensive nature, code review within team gradually increases the time and resources spent on review activities as projects become more complex and extensive~\cite{rigby2013convergent,rigby2014peer}. Therefore, beside the popularity of team code review~\cite{johnson1998reengineering}, modern code review has evolved from a formal and rigorously regulated process~\cite{sadowski2018modern,fagan2011design} to a less stringent practice~\cite{baum2017choice,baum2016faceted,rigby2013convergent}.

\textcolor{black}{\textbf{Public Code Review} (PCR) is a type of modern code review task developed in the Software Question Answering (SQA) community, which helps developers explore efficient and high-quality review services. Unlike traditional code review, PCR is typically directed towards a broader developer community, where users can submit review requests without being tied to a specific project context. This openness enables PCR to gather diverse perspectives and expertise on code, but the lack of project structure often leads to submissions that lack context or clear background information, introducing challenges in management and quality control of code review. Existing research on code review has already recognized the importance of quality assurance mechanisms~\cite{baum2017optimal,baum2016need,sadowski2018modern}. Therefore, this issue should also be given attention in public code review tasks. In open environments lacking project structure, appropriate quality control not only helps guide users to submit clearer and more standardized review requests, but also provides safeguards for identifying potential issues and facilitating efficient collaboration.}

\begin{figure}[h]
  \centering
  \includegraphics[width=0.7\linewidth]{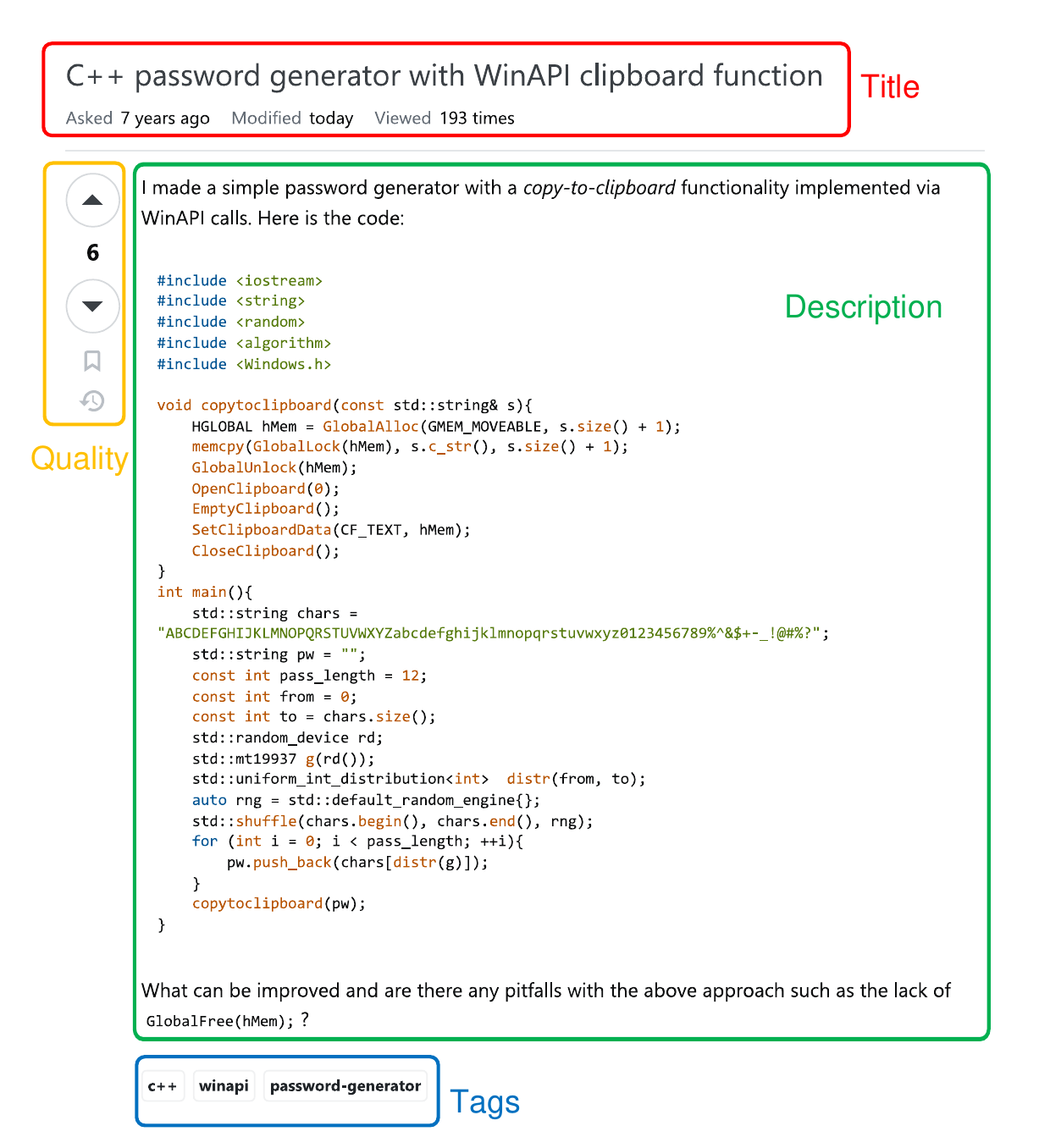}
  \caption{An Example of a Public Code Review Request.}
  \Description{the request body, tags, and request contributions}
  \label{fig:fig1}
\end{figure}

\textcolor{black}{In the PCR process implemented by the SQA community, request quality assurance for code review service mainly depends on developers. Developers are required and encouraged to submit PCR requests in the community. A review request\footnote{https://codereview.stackexchange.com/questions/284179/proper-implementation-of-signal-handler-and-multithreading-pthread} consists of several parts, as shown in Fig.~\ref{fig:fig1}, including \textit{title}, \textit{request description (text and code snippets)}, \textit{quality}, and \textit{tags}. 
The necessity of these code review requests is evaluated by public via \textbf{request necessity prediction}~\cite{chen2024unified} subtask after the requests are posted. Meanwhile, developers always need to choose technical terms as review tags. These tags can be used to match requests with suitable reviewers via \textbf{tag recommendation}~\cite{li2023dual,47} subtask. Since the quality and responsiveness of public code reviews typically rely on manual efforts by developers, which can be labor-intensive, the demand for transitioning from manual reviews to intelligent services has gradually emerged. }

Existing methods on PCR are mostly based on pre-trained language models. However, these models typically exhibit generalized capabilities, and their understanding of domain-specific knowledge in the PCR field, such as code-related semantic patterns, still has room for improvement. Early work regarded code snippets as non-linguistic characters, which impeded the model’s performance in understanding technical content. For example, the TagDC method proposed by Li et al.~\cite{24} and the PROFIT model proposed by Nie et al.~\cite{27} treated code snippets as irrelevant information and discarded them altogether. Recent studies have discovered valuable semantic information embedded in code snippets, but these studies mostly employed the same processing methods as they do with text~\cite{xu2021post2vec,29}. Knowledge enhancement can effectively address the gaps in existing models when handling programming terms and error patterns. For instance, entries on algorithms, programming languages, and common programming errors can help the model better understand the functionality and potential issues of complex code snippets, thus improving its performance in PCR tasks.

To this end, we propose \underline{K}nowledge-guided \underline{P}rompt learning for \underline{P}ublic \underline{C}ode \underline{R}eview (KP-PCR). Specifically, we first design a task-descriptive prompt template that restructures the request necessity prediction subtask and the tag recommendation subtask into a generative task through a knowledge-guided prompt template, and input the request body (including tags and descriptions) into the prompt construction module of a language model. We reformulate both subtasks via \textbf{1) text prompt tuning} which converts two subtasks into Masked Language Model (MLM) by constructing prompt templates using hard prompt; and {2) knowledge and code prefix tuning} which introduces knowledge guidance from fine-tuned DeepSeek by soft prompt, and uses program dependence graph to characterize code snippets. Finally, the request necessity prediction and recommendation tags output predicted results through an answer engineering module.
In addition, we delve deeper into the time complexity analysis of our KP-PCR with lightweight prefix based operations to introducing knowledge, which reveals that our KP-PCR can improve task accuracy performance without compromising overall efficiency. 
Experimental results demonstrate that our KP-PCR outperforms baselines in terms of accuracy-related metrics across both subtasks. 
Furthermore, we explore the impact of prompt learning construction and knowledge guidance sources on the results, and use a case study to compare our KP-PCR with Llama3, CodeLlama and DeepSeek.

The main \textbf{contributions} of this study can be summarized as follows:
\begin{itemize}
\item From the perspective of developers, to satisfy the request quality assurance, a unified method, called KP-PCR, is proposed for the request necessity prediction subtask and the tag recommendation subtask. 
\item We introduce knowledge guidance from fine-tuned DeepSeek to enhance the model's semantic understanding of domain knowledge in developers' PCR requests. 
\item Experimental results on the PCR dataset for the time span 2011-2023 demonstrate that our KP-PCR outperforms the baselines of two request quality assurance subtasks. 
\end{itemize}

This paper extends the preliminary version of our study, as presented in~\cite{chen2024unified}. In comparison, this extended version (1) introduces knowledge guidance from fine-tuned DeepSeek to generate prefix vectors as soft prompt for the representation of code snippets, (2) changes the code parsing mode to PDG to get more control dependency information, (3) analyzes the time complexity to prove that our KP-PCR can improve the accuracy of request necessity prediction and tag recommendation while maintaining the overall efficiency of our KP-PCR method, (4) comprehensively evaluates the effects of our task-descriptive prompt template to explore the influence of knowledge guidance and code parsing on unifiedly modelling the above two subtasks, and (5) conducts a case study to compare our KP-PCR with Llama3, CodeLlama and DeepSeek. Additionally, we have examined the challenges of how to design hard and soft prompt templates and delved further into the originally explored unified modelling in the initial study. More details on the extension points compared to our preliminary study in~\cite{chen2024unified} can be found in Section~\ref{sec:7.5}. 

The remainder of this paper is structured as follows. Section~\ref{sec:2} presents a motivating example, and Section~\ref{sec:3} introduces the task addressed in this paper. Section~\ref{sec:4} details the proposed method. Section~\ref{sec:5} provides the experimental setup, and Section~\ref{sec:6} discusses the experiment results and presents a case study to explore the utilization of knowledge guidance by our KP-PCR. Section~\ref{sec:7} describes the related work. Section~\ref{sec:8} concludes this work with future directions.

\section{Motivating Example}\label{sec:2}

\textcolor{black}{Public Code Review (PCR) differs from internal team code reviews in its service format, as the quality of the PCR process primarily depends on the developers who write and submit review requests, rather than on the reviewers themselves. While existing research has been effective in improving review quality, there remains significant room for improvement. One of the limitations of the current approach is that pre-trained language models still require improvement in their ability to learn domain-specific knowledge. The text of review requests may also include vocabulary that is absent from the language model’s training data. During tokenization, the model may split these out-of-vocabulary words into subwords or individual characters, which limits its ability to accurately process the text and complete PCR tasks.} 

Table~\ref{tab:1} provides an example from the tag recommendation subtask, in which the review request includes the abbreviation ``OOP'' to describe a concept. However, this term is not covered by the language model, and the model may struggle to connect ``OOP'' with its full form, ``Object Oriented Programming''. In contrast, knowledge guidance provides definitions that link these terms, helping to resolve such discrepancies. 

To address this limitation, we propose fine-tuning a large pre-trained model, such as DeepSeek. By fine-tuning DeepSeek-7B on Wikipedia's extensive collection of programming terminology, technical definitions, and domain-specific concepts, we can improve the model's ability to understand and generate code-related content more accurately. This approach allows the model to internalize valuable knowledge during the training process, ensuring that the learned domain-specific knowledge is embedded directly within the model's parameters.

\begin{table*}
\small
\captionsetup{}
  \caption{Example of Professional Knowledge Guidance}
  \label{tab:1}
  \begin{tabular}{>{\centering\arraybackslash}p{5cm}c>{\centering\arraybackslash}p{5cm}}
    \toprule
     Data& Tag&Knowledge Guidance\\
     \midrule
An online store, being converted from procedural to {\bfseries OOPI} just start to learn 
{\bfseries OOP}, and it's far more interesting than procedural style.I have a complete working online store written in procedural 
style...
& {\bfseries object oriented}
& {\bfseries Object-oriented programming (OOP)} is a programming paradigm based on the concept of 
objects, which ...\\
    \bottomrule
  \end{tabular}
\end{table*}

\section{Task Description}\label{sec:3}

In the PCR process conducted by the SQA community, the quality assurance of requests primarily relies on developers. Developers are required to submit necessary PCR requests to the community, and the necessity of these requests is assessed publicly via the \textbf{request necessity prediction} subtask after submission. At the same time, developers need to select several technical terms as review tags, which can be used to match requests with appropriate reviewers through the \textbf{tag recommendation} subtask.

In this section, we will first describe a public code review process, and then outline how these request quality assurance subtasks (request necessity prediction and tag recommendation) serve the public code review process. This process consists of five core steps (as shown in Fig.~\ref{fig:task_2}), starting from the submission of changes or optimization requests by the code developer in Step \textcircled{1}, to the submission of reviews by reviewers in Step \textcircled{5}.

\subsection{Public Code Review Process}
\begin{figure}[h]
  \centering
  \includegraphics[width=\linewidth]{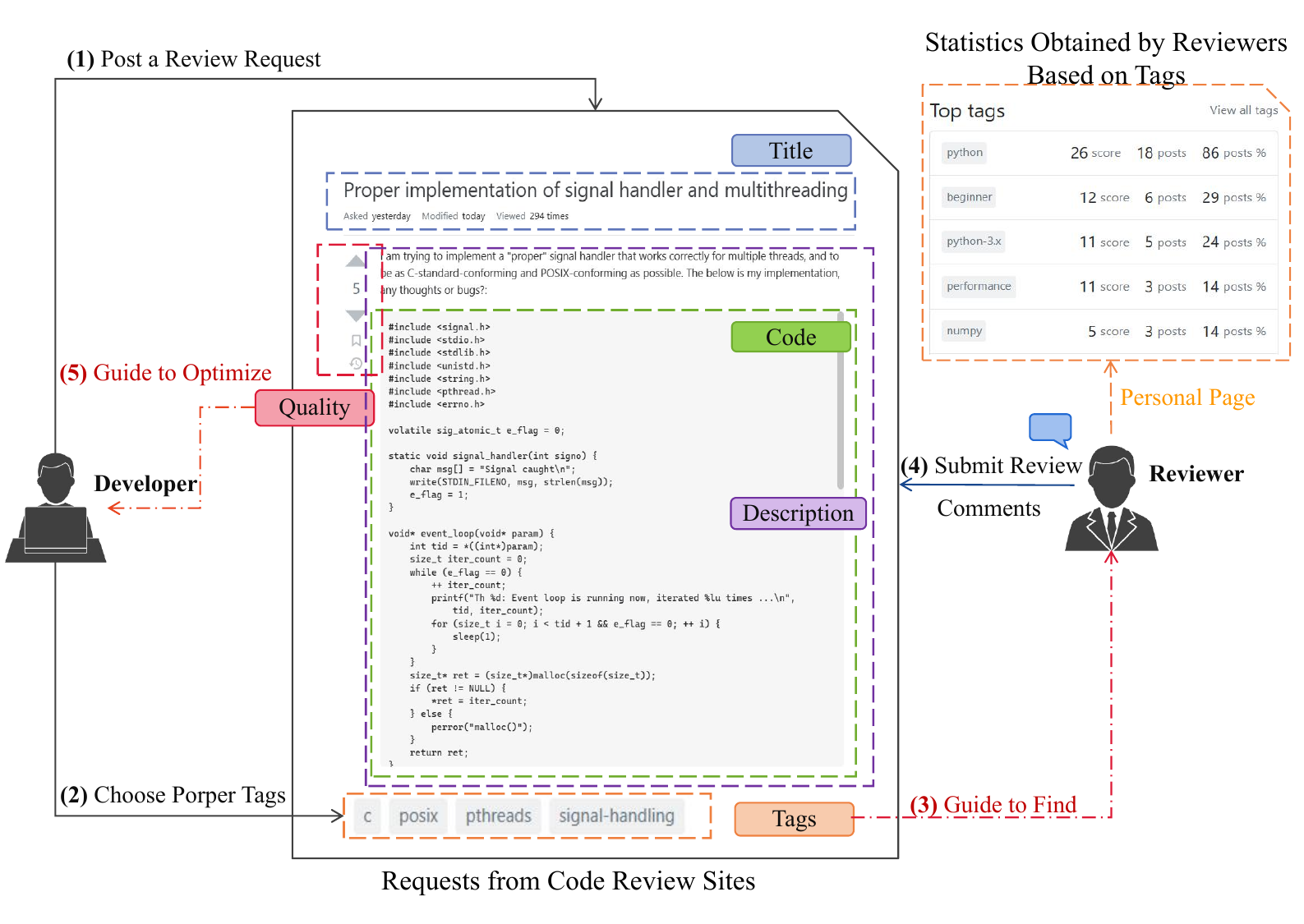}
  \caption{An Example of a Public Code Review Process.}
  \Description{a review process can be abstracted into 5 step}
  \label{fig:task_2}
\end{figure}

First, the public code review process will be introduced through an example of public code review. As shown in Fig.~\ref{fig:task_2}, a review process can be abstracted into 5 steps, labeled as \textcircled{1}-\textcircled{5}. Code review typically involves two roles: \textbf{the developer who submits the code request and the reviewer who provides suggestions and comments}. In Step \textcircled{1}, the developer writes and submits a review request. In Step \textcircled{2}, practitioners in the community review and evaluate the requests to allocate necessary feedback. Subsequently, in Step \textcircled{3}, developers are asked to select several relevant technical terms to accompany their request. These tags are then used in Step \textcircled{4} to match the request with appropriate reviewers, who provide their feedback on the code in Step \textcircled{5}.

In a review request, there are often many changes and optimizations needed in different parts of the code. Reviewers need to spend a significant amount of time reviewing all the changes. However, it is often the case that most changes are minor and do not require comments or suggestions. Li et al.~\cite{li2022automating} initially considered the task of assessing code change quality in the internal team code review process, which involves predicting whether a code change is of high quality and ready to be accepted in the review process. This work enhances the efficiency and quality of the review process by pre-evaluating before the request is reviewed. Therefore, transitioning to the public code review process, there is a task need for predicting request necessity. Request necessity prediction learns from the existing conventions of public code review and community feedback to determine the standards for submitting review requests, thereby ensuring review quality and reducing the time cost of unnecessary requests.

The task of tag recommendation is common in the software engineering community, typically used for labeling requests and enhancing visibility. In public code review, reviewers browse request content and conduct evaluations based on these tags. Therefore, appropriate tag recommendations contribute to improving review efficiency.

Overall, our task logic will focus on two processes initiated by developers, aiming to optimize the services of public code review. This includes integrating the task of predicting request necessity with Step \textcircled{2}, to provide necessary feedback to practitioners in advance. This can guide them to optimize request expressions and improve request visibility while reducing the time spent by reviewers in Step \textcircled{5}. Additionally, the tag recommendation task is associated with Step \textcircled{4}, allowing requests to be correctly matched with appropriate reviewers based on the usage of tags in the PCR community.

\subsection{Request Quality Assurance Task Definition}
\label{sec:3.2}

The effectiveness of the PCR process depends on the interaction between developers and reviewers. However, this interaction can be time-consuming due to the need to identify issues, find suitable reviewers, etc. While current methods focus on reducing the time cost for reviewers, they overlook the impact of developers on the review process. For the PCR process, code review requests that comply with review standards are crucial for high-quality review responses and a conducive community atmosphere. To address this issue, this study proposes a quality assurance service for public code review, mainly relying on the request necessity prediction subtask and the tag recommendation subtask.

When considering a request in public code review, it is assumed that there exists a corpus of public code review requests (denoted as $R$), a set of tag labels (denoted by $L$), and a set of quality labels for all PCR requests (denoted by $Q$).  Typically, given a public code review request ${req}_{i} \in {R}$, the request consists of a title ${title}_{i} \in {T}$, and a description ${des}_{i} \in {D}$, where the description ${des}_{i}$ includes both text ${tex}_{i}$ and code snippets ${code}_{i}$. 

\textbf{(1) The subtask of request necessity prediction.} The objective of the request necessity prediction is to obtain the function quality that maps $r$ to the necessity label ${q} \in {Q}$. A public code review request ${req}_{i}$ corresponds to a necessity label ${q}_{r}$. Therefore, the necessity prediction subtask can be regarded as a classification problem. The goal is to determine an optimal function $f$ that maps the necessity label $q$ to a label similar to the actual necessity label ${q}_{r}$.

\textbf{(2) The subtask of tag recommendation.} The objective of the tag recommendation subtask is to obtain a function that maps ${req}_{i}$ to a set of labels 
$l=\left\{ l_1,l_2,\ldots ,l_m \right\} \subset L$ 
that are most relevant to the request $r$. A review request 
$req_i$ 
corresponds to several labels, where 
$l_r=\left\{ l_{r1},l_{r2},\ldots ,l_{rk} \right\} \subset l$, 
$k$ 
represents the number of true labels of the review request. Therefore, the tag recommendation subtask can be viewed as a multi-label classification problem. The goal is to identify an optimal function
$f_{tag_rec}$ 
that maps the label subset $l$ to the true label set $l_r$ as closely as possible.

\textbf{(3) Unified task definition.} We design a descriptive prompt template for the two subtasks, summarized as: $T\left( \cdot \right) =``x\left[ MASK \right]$'' (where $x$ represents a text string). The input sequence $r$ of the request is passed through this descriptive prompt and refactored into $r_{prompt}=T\left( r \right)$. The request necessity prediction and tag recommendation subtasks are made by filling in the position of $\left[ MASK \right] $. Our unified task is how to improve the proximity of the predicted filler words against the truth label. We denote the total number of training examples as $N$, the total number of available labels as $M$, and the number of labels in the training data as $l$. The number of $MASK$ is determined by the maximum length of the labels in the two subtasks.

\section{KP-PCR Method}
\label{sec:4}

In this section, we propose a knowledge-guided prompt learning for PCR called KP-PCR to address the issue of insufficient learning of professional knowledge in public code review.

\subsection{Framework Overview}

\begin{figure}[h]
  \centering
  \includegraphics[width=\linewidth]{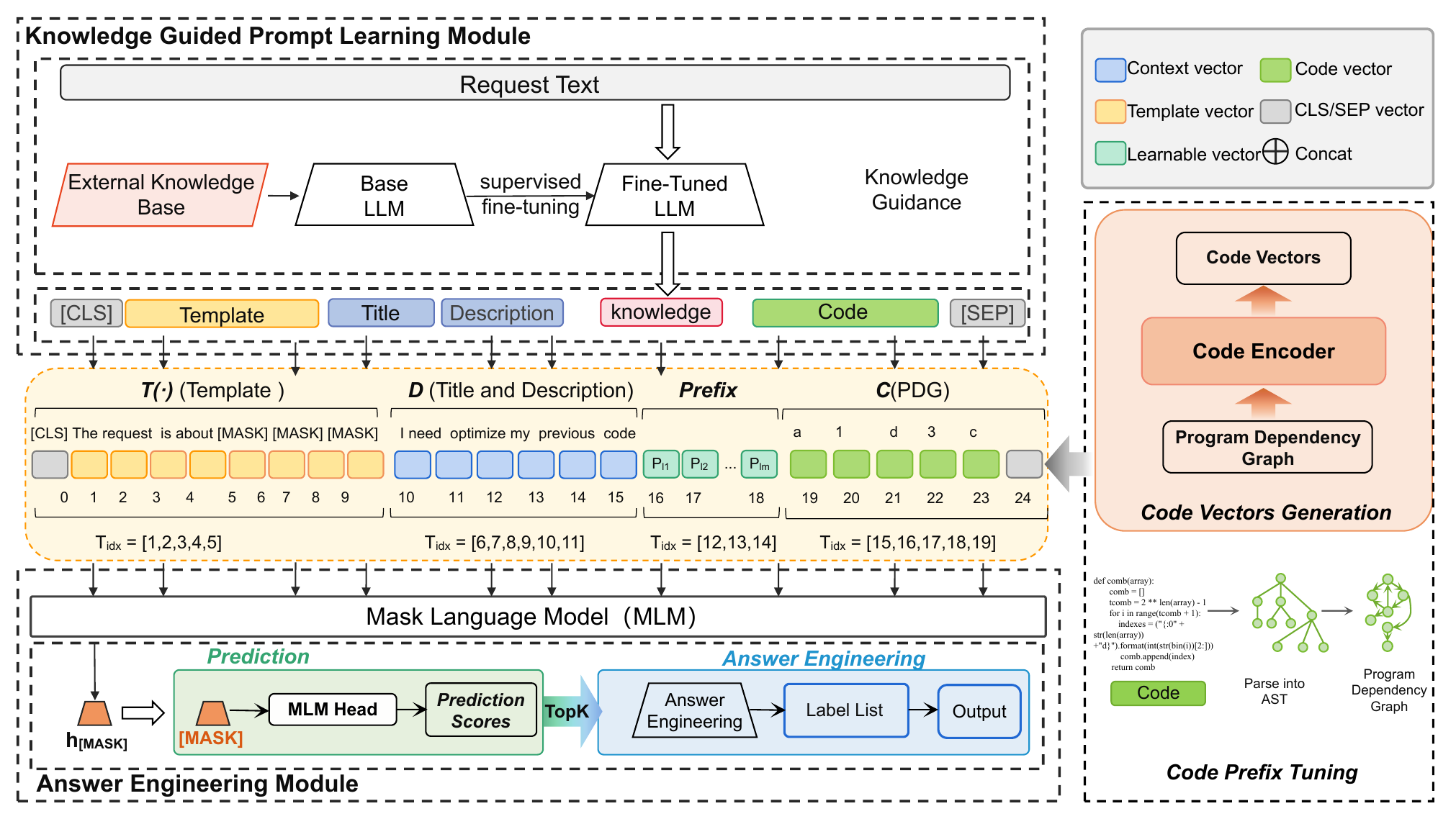}
  \captionsetup{labelfont={color=black}}
  \caption{\textcolor{black}{The Framework of KP-PCR.}}
  \Description{Includes three parts of prompt tuning architecture: text prompt tuning, code prefix tuning, and unified prompt tuning.}
  \label{fig:framework}
\end{figure}

Fig.~\ref{fig:framework} illustrates the overall framework of KP-PCR which has two modules, knowledge-guided prompt learning module and answer engineering module. 

The knowledge-guided prompt learning module consists of text prompt, code prompt and knowledge-guided prompt. First, the request text is input into prompt learning module. The text prompt reconstructs the request text into a generative task, and the code prompt refactors the code snippets in text into the form of program dependence graph. These processes are described in more detail in Section~\ref{sec:4.2}. The request is then fed into the fine-tuned DeepSeek. The text generated by DeepSeek is used as knowledge guidance, which is then fed into the model in the form of prefix vectors. Further, the obtained prompt template is input into the Masked Language Model (MLM) to get the predicted word list. Then the answer engineering module maps the predicted words to the label list to get the final labels.

\subsection{Unified Prompt Template Design}
\label{sec:4.2}

\subsubsection{Text Prompt}

This subsection elaborates on the process of restructuring subtasks through task descriptive prompt templates, which involves the following two steps:

{\bfseries (1) Task Descriptive Prompt Template}

First, a descriptive prompt template $T\left( \cdot \right) $ as a hard prompt is applied to map the input request body $req_I=\left\{ title_i,\ des_i \right\} $ to prompts. The prompt template retains the original tokens from $req_i $ and includes at least one [MASK] for PLM to fill in label words for both subtasks. Specifically, this study modifies the descriptive prompt text based on the subtask objectives to reconstruct the data for different tasks. 

Then, by using the descriptive prompt: ``\textit{The requested label is [MASK][MASK][MASK]. Is this request necessary for review? [MASK]}'', the tasks of necessity prediction and tag recommendation are restructured within the same framework.

Finally, with a guiding prompt ``Request: '', the model is directed to explicitly recognize the request content, and concatenate the title $title $, description $des $ (including text content $text $ and code snippet content $code $), and the template $T\left( \cdot \right) $ together. After concatenation, we obtain a text input $req_i $ for the model, as shown in Equation~\ref{eq:input}:

\begin{equation}
Input_i=\left\{ T\left( \cdot \right) ,title_i,des_i \right\} 
\label{eq:input}
\end{equation}

{\bfseries (2) Request Text Encoding}

$\left[ CLS \right] $ and $\left[ SEP \right] $ are special tokens used in BERT-based pre-trained language models. $\left[ CLS \right] $ is used to mark the beginning of a sentence, and the representation vector obtained from the encoder can be used for subsequent downstream tasks. $\left[ SEP \right] $ is used to mark the end of a sentence or to separate two sentences. The input sequences are fed into a tokenizer which identifies each input with $\left[ CLS \right] $ and $\left[ SEP \right] $.

Next, as shown in Equation~\ref{eq:input_2}, the complete input sequence obtained earlier is fed into the model's embedding layer for encoding.

\begin{equation}
Input_i=\left\{ \left[ CLS \right] ,T\left( \cdot \right) ,title_i,des_i,\left[ ESP \right] \right\} 
\label{eq:input_2}
\end{equation}

The embedding layer of the encoder model initializes each token according to the tokenizer's tokenization results. Based on pre-learned vector representations, a representation vector for the input sequence is obtained.

\subsubsection{Code Prompt}

This subsection systematically elaborates on the methodology for constructing prefix vectors based on program dependence analysis. 

\begin{figure}[h]
  \centering
  \includegraphics[width=\linewidth]{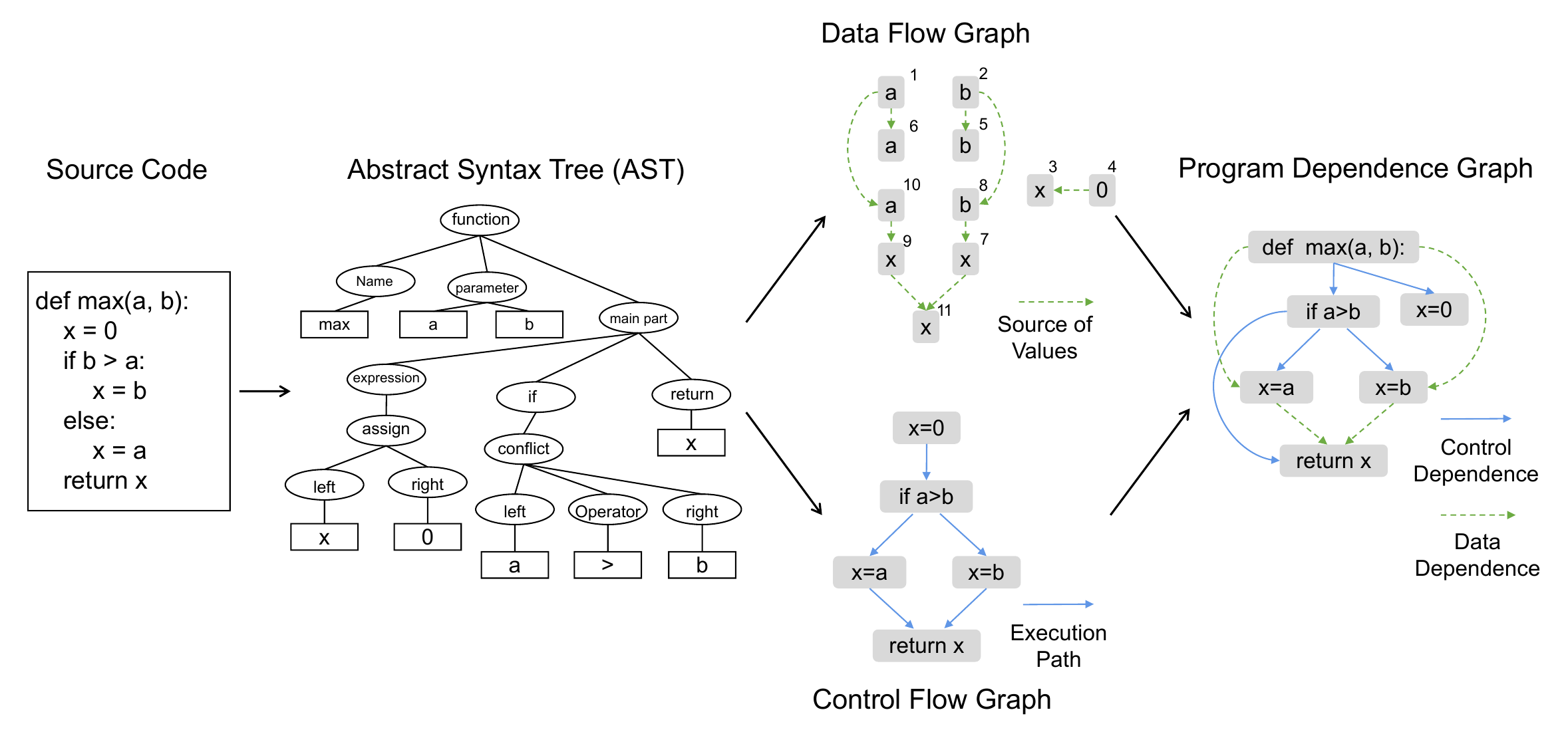}
  \caption{A Code Snippet Processing Example.}
  \Description{the language of the code snippet is identified.the abstract syntax tree is further parsed into the form of a PDG}
  \label{fig:pdg}
\end{figure}

In the label recommendation task, labels are often closely related to the behavior of code under different control flow paths. Similarly, in the necessity prediction task, it is often essential to understand both the control logic and the data dependencies within the surrounding context. Program Dependence Graph (PDG) captures both control dependencies and data dependencies, allowing for a comprehensive representation of the program execution logic. By hierarchically parsing code Data Flow Graph (DFG) and Control Flow Graph (CFG), as shown in Figure~\ref{fig:pdg}, we ultimately fuse these structures to build a Program Dependence Graph (PDG), enabling formal semantic representation of code. 

Firstly, the language of the code snippet is identified, as each programming language has different coding styles and syntax. Therefore, it is necessary to parse the code snippet for each language, extract important nodes and variable information in the form of an abstract syntax tree (AST) after obtaining the language category. Secondly, the abstract syntax tree is further parsed into the form of DFG and CFG. When converting incomplete code snippets, technique such as filling in gaps with placeholder nodes is used to handle incomplete connections. 

For a code snippet $C$, its DFG is defined as a directed graph $G_{dfg}=(V_{dfg}, E_{dfg})$. Node set $V_{dfg} = \{v | v \in \text{Variables}(C) \cup \text{Operations}(C)\}$ represents variables and operators. Edge set $E_{dfg}=\{\left( v_i,v_j \right) |v_j\rightarrow_{dfg} v_i\}$, where $v_j\rightarrow v_i$ denotes a data dependence from $v_i$ to $v_j$. 

The CFG of code snippet $C$ is defined as a directed graph $G_{cfg}=(V_{cfg}, E_{cfg})$. Node set $V_{cfg} = \{b | b \in \text{BasicBlocks}(C)\}$, where basic blocks are maximal sequences of branch-free instructions. Edge set $E_{cfg} = \{(b_i, b_j) | b_i \rightarrow_{cfg} b_j \}$, where $b_i \rightarrow_{cfg} b_j$ denotes immediate control flow from $b_i$ to $b_j$.

Finally, duplicate nodes between the DFG and CFG are merged to generate PDG with comprehensive data and control dependency representation.

\subsubsection{Knowledge-Guided Prompt}

In order to optimize the code representation and better guide the knowledge embedded in the request text and code snippets, we use the continuous prompting method (soft prompt)~\cite{liu2023pre} to introduce knowledge guidance. Specifically, we first extract the professional term from the request using string matching, then initialize the $P_{prefix}$ vector prefix by concatenating the knowledge interpretation representation and concatenate the representation vector of the program dependence graph of the code snippet after this prefix vector. Finally, the whole vector is represented as the model code snippet instead of the original code snippet. Further details on these extensions can be found in Section~\ref{sec:4.3}.

\subsection{Knowledge-Guided Prompt Learning Module}
\label{sec:4.3}

\subsubsection{Knowledge-Guided Prompt Learning}
\label{sec:4.3.1}

Soft prompting is one of the prompting paradigms, which operates in the embedding space of the model and adjusts the prompt parameters for downstream tasks. Currently, this method has shown superior performance in other data not in natural language, such as image encoding and visual question answering~\cite{liu2023descriptive,tsimpoukelli2021multimodal}. This indicates that soft prompts are well-suited for learning and optimizing information with such graph structures, owing to the ability of vector embeddings in the embedding space to better adapt to non-natural language structures. The PCR task is not merely about code quality detection or generation; it involves a multi-step code review process, which includes predicting whether the code needs review and how to select appropriate labels for the code review task. Knowledge-guided learning in PCR can help the model understand the context of the code and identify common algorithms, functions, or libraries involved in the code snippets, thereby enhancing the model’s comprehension and judgment of code review requests. 

\textcolor{black}{To enhance the model's performance by integrating external knowledge, our framework uses LLM, specifically introducing the DeepSeek-7B. The reason for choosing this model is that it has demonstrated proven adaptability in code understanding tasks, and its parameter scale is relatively small, requiring less computational resources. It can be deployed in an environment with at least 16GB of GPU memory. By performing label retrieval on Wikipedia for the dataset, we have obtained 1,000 articles related to software engineering, programming languages, algorithms, and technical concepts. These articles are then preprocessed to ensure that their format was suitable for the LLM's input requirements. The data was cleaned to remove irrelevant content, leaving only technical definitions, code examples, and domain-specific knowledge. DeepSeek-7B is then fine-tuned using a supervised learning approach, with a learning rate of $2\times 10^{-5}$, a batch size of 8, and trained for 10 epochs. Regarding the consumption of inference time, after testing under the condition of a 512-token input, the response time for every 1000 rounds is 8 minutes, and the average response time per round is 480 milliseconds. The additional computational overhead brought by using LLM is reasonable and controllable.}

This study uses the labels of specialized terminology as a benchmark and selects words that are not present in the vocabulary of the pre-trained language model. First, string operations are used to split words based on spaces. Additionally, the split words undergo fuzzy case, tense, and part-of-speech transformations. Next, the processed text strings are used for string matching to extract domain-specific knowledge words contained in each request body, which are then input into the fine-tuned DeepSeek model to obtain their expert definitions. Finally, the definitions of domain-specific knowledge words are concatenated to form their representations as input into the model. And they work in the form of prefix vectors to aid in learning the knowledge within the request body.

Here, taking an input request $req_i$ as an example, first, based on the previous data processing, we can obtain two parts of a PCR request: 

1. Natural language text $text_i$, composed of the title and description (text part) of the request; 

2. Code snippet $code_i$, representing the code snippet extracted from the description. 

\renewcommand{\algorithmicrequire}{\textbf{Input:}}
\renewcommand{\algorithmicensure}{\textbf{Output:}}

\begin{algorithm}[h]
  \caption{Knowledge-Guided Prompt Learning}
  \label{alg::conjugateGradient}
  \begin{algorithmic}[1]
    \Require
      $req=\left( title,\ text,\ code \right)$
    \Ensure
      $r_{\widetilde{prompt}}$
 
    \State $Graph_c\gets Graph\left( code \right)$, $Graph\left( code \right) $ represents the program dependence graph of the code snippet;       
    \State $h_{Graph_c}\gets PLM\left( Graph_c \right)$, $h_{Graph_c}$ represents the code hidden vector obtained through the encoding model;
    \State $knowledge=req\cap \text{wiki}$;
    \State $P_{prefix}=PLM\left( knowledge \right)  ,\ PLM\left( knowledge \right) $ represents the representation vector obtained through the encoding model;
    \State $r_{\widetilde{prompt}}=\left\{ T\left( \cdot \right) ,\ \widetilde{title},\ \widetilde{text},\ P_{prefix},\ h_{Graph_c} \right\} $;

    \State return $r_{\widetilde{prompt}}$
  \end{algorithmic}
\end{algorithm}

The method for processing the code snippet and constructing the prefix vector is shown in Algorithm 1. In Step \textcircled{1}, a code snippet $code_i$ is parsed into a program dependence graph to represent the dependency relationship among code variables (as shown on the right side of Fig.~\ref{fig:dataflow}). 
Then, in Step \textcircled{2}, the variable relationship graph is input into a pre-trained code encoding model for code tasks to obtain the hidden vector of the code snippet. In this work, the GraphCodeBERT~\cite{63} model is used as the encoding model.
In Step \textcircled{3}, professional knowledge terms are extracted from the request using string matching.
Step \textcircled{4} defines a prefix vector $P_{prefix}$, then initializes the prefix vector $P_{prefix}$ with the representation vector of knowledge processed by the encoding model. 
Finally, the entire vector replaces the original code snippet as the model's code representation. 

Here, $P_{idx}$ is used to represent the sequence of prefix indices, and $\sigma $ is used to represent the non-linear output obtained from the hidden layers of the code pre-trained language model. To address the issue of low performance and high variance caused by the random initialization of $P_{prefix}$, we initialize the prefix vector $P_{prefix}$ with the text of knowledge guidance and then use an immutable program dependence graph based on semantic relationships to represent the meaning of individual code snippets. This allows us to guide the model to learn professional knowledge from the request text through knowledge guidance and to globally and locally understand the meaning of each code snippet through the form of prefix vectors. 

In order to measure the efficiency of Algorithm 1, we analyze its time complexity. To this end, we consider the computational overhead in each step.
In Step \textcircled{1}, the size of a code snippet determines the time complexity to convert code into a program dependence graph. If there are $n$ variables or code fragments in the code snippet, generating a program dependence graph requires checking the dependencies between each pair of variables, and consequently the time complexity is $O\left(n^2\right)$.
Then, in Step \textcircled{2}, the time complexity of GraphCodeBERT generally depends on the architecture of its pre-trained model and the input graph size of a program dependence graph. Overall, the complexity of Step 2 can be seen as $O\left( V+E \right)$, in which $V$ is the number of nodes in the graph, and $E$ is the number of edges. The complexity of this step is also $O\left(n^2\right)$, because the encoding process is proportional to the size of the graph.
Step \textcircled{3} uses string matching whose time complexity is $O\left( s \right)$, in which $s$ is the length of the request text. 
The time complexity of Step \textcircled{4} depends on the size of the encoding model and the length of the knowledge representation vector. The process of initializing a vector can be regarded as a constant time operation, and the complexity is $O\left( d \right)$, in which $d$ is the length of the vector. 
Therefore, the overall time complexity of the algorithm can be expressed as $O\left(n^2+s+d\right)$. Since the complexity of language model representation is much higher than that of string matching, i.e., $O\left(n^2\right)\gg O\left(s \right)$ and $O\left(n^2\right)\gg O\left(d \right)$, and the time complexity can also be expressed as $O\left(n^2\right)$. 

Our preliminary study~\cite{chen2024unified}, where the time complexity is also $O\left(n^2\right)$, has an actual runtime of approximately 8 hours per epoch. The actual runtime of each epoch for our KP-PCR method is about 8 hours and 20 minutes, which is nearly the same. However, it can be found that according to the experimental results in Section~\ref{sec:6}, our KP-PCR method can improve the task accuracy of request necessity prediction and tag recommendation.

\subsubsection{Loss Function}

\begin{figure}[h]
  \centering
  \includegraphics[width=\linewidth]{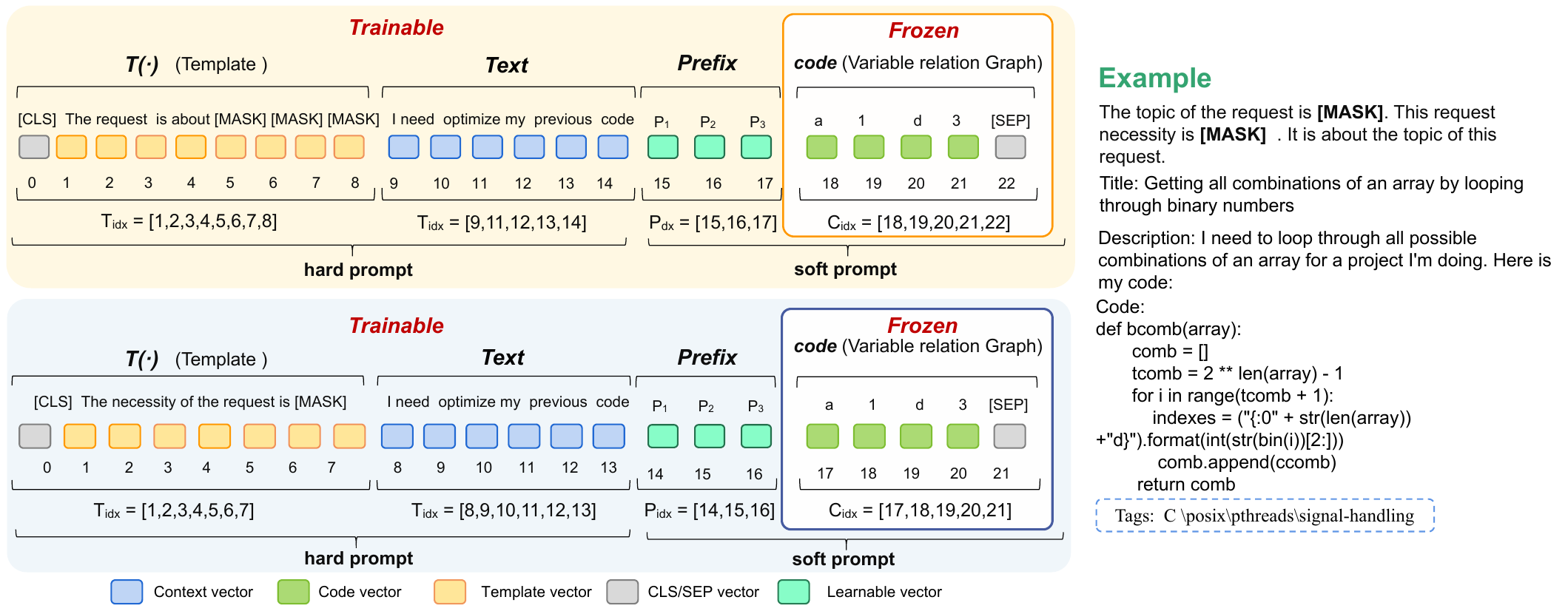}
  \caption{Trainable and Frozen Parameters.}
  \Description{concatenate the obtained text segment vector and code snippet vector into the input prompt of the Masked Language Model (MLM).}
  \label{fig:prefix}
\end{figure}

Fig.~\ref{fig:prefix} formalizes the construction of task inputs through an example request, showcasing the joint construction of task description prompt templates (hard prompt) and prefix vectors (soft prompt). It is divided into trainable vector parts and frozen vector parts that are not involved in training.

First, the obtained text segment vector and code snippet vector are concatenated into the input prompt of the Masked Language Model (MLM). Within the entire input $\widetilde{c}$ is an immutable vector, while the rest of the vectors are updated based on the gradient optimization target. As shown in Fig.~\ref{fig:prefix}, the vector $\widetilde{c}=h_{Graphc}$ is encoded by the hidden layer of the pre-trained code language model is frozen and not updated, whereas the prefix is initialized using the code snippet text. Finally, the vector $h_{PREFIX_i}$ can be trained. The dimensions of the code prefix tuning initialization trainable matrix $P_{\sigma}$ (parameterized by $\sigma$) are $|Pidx|\times \text{dim}\left( h_{Xprefixi} \right) $ to store the prefix parameters.

Since multiple [MASK] tokens may appear in the descriptive prompt template, all masked positions are considered for prediction. The final learning objective of our KP-PCR framework is to consider multiple [MASK] tokens in a descriptive prompt template, using the sum of losses from all [MASK] positions to optimize the model. Specifically, the final learning objective of KP-PCR is to maximize Equation~\ref{eq:logp}:

\begin{equation}
\log p_{\phi}\left( Lable_{pre}|Req \right) =\prod\limits_{j=1}^n{p_{\phi}\left( \left[ MASK \right] _j=\varphi _j\left( label_r \right) |T\left( req \right) \right)}, 
\label{eq:logp}
\end{equation}

where $\varphi$ is the predicted probability distribution, and $\varphi _j\left( label_r \right) $ represents the mapping from the $j$-th $h_{MASK}$ token to the requested label word $label$ in the input sequence. $\phi$ denotes the parameters of the language model that are updatable. $n$ is the number of masked positions in the descriptive prompt template $T\left( req \right) $.

\subsection{Answer Engineering Module}

After training the model with Equation~\ref{eq:logp}, the final labels need to be obtained, so the answer engineering module maps the predicted words to the label list. The answer engineering module maps the $TOP K$ token list $pre_{list_i}$ to the predicted labels $f_{ans}\left( pre_{list_i} \right)$ through a mapping function $f_{ans}$. The answer engineering function $f_{ans}$ maps the predicted words to the final label set $Label$. Finally, the mapped answer $f_{ans}\left( pre_{list_i} \right)$ is the final top-k outputs $v_i$. The mapping function of the answer engineering first compares the predicted words with the label word list to find the complete label by predicting partial label words. Secondly, for data where the predicted words do not exist in the entire label word list, the mapping is completed using the minimum edit distance algorithm. Edit distance, also known as Levenshtein distance, refers to the minimum number of editing operations required to transform one string into another. For example, for the strings ``if'' and ``iff'', you can achieve the transformation by inserting or deleting a ``f''.

In general, the smaller the edit distance between two strings, the more similar they are. If two strings are equal, their edit distance is 0 (no operations are needed). In this study, edit distance is used to evaluate the distance between predicted words and true words. If a predicted word is not in the list of true word labels, then by calculating the edit distance between the predicted word and all true words, we can find the true word with the minimum edit distance to the predicted word, thus correcting the predicted word.

\section{Experimental Setup}
\label{sec:5}

\subsection{Research Questions}

Our research goal is to provide quality assurance services for PCR from the perspective of developers. Therefore, our method is evaluated from the following three questions.

\begin{tcolorbox}[arc=0mm,width=\columnwidth,
                  top=0mm,left=0mm,  right=0mm, bottom=0mm,
                  boxrule=.75pt]
\textbf{RQ1}: Can our KP-PCR effectively deal with the request necessity prediction subtask and the tag recommendation subtask in a unified modelling?
\end{tcolorbox}

Our method and SOTA models are applied to examine necessity prediction and tag recommendation subtasks respectively, and their performance on the two subtasks is evaluated.

\begin{tcolorbox}[arc=0mm,width=\columnwidth,
                  top=0mm,left=0mm,  right=0mm, bottom=0mm,
                  boxrule=.75pt]
\textbf{RQ2}: Is our knowledge-guided prompt learning effective for PCR?
\end{tcolorbox}

Different prompt templates are explored to know the influence of propmt learning on model performance under different designs.

\begin{tcolorbox}[arc=0mm,width=\columnwidth,
                  top=0mm,left=0mm,  right=0mm, bottom=0mm,
                  boxrule=.75pt]
\textbf{RQ3}: How does our unified prompt template perform compared large language models?
\end{tcolorbox}

we fine-tuned Llama3, CodeLlama and DeepSeek using the entire training dataset, and then evaluated them on the test set, comparing their performance with our model.  

\subsection{Dataset and Metrics}

\subsubsection{Dataset Description}

\textcolor{black}{To evaluate the effectiveness of our KP-PCR, experimental data is from the Code Review site within Stack Exchange, a reputable and well-known information-sharing platform. The selection of the dataset follows the criteria of being open, authoritative, and current. Stack Exchange is a large network of question and answer (Q\&A) websites covering a wide range of topics, including but not limited to software domains such as programming, Linux \& Unix, and code review. Each website within this network focuses on a specific theme, and the questions, answers, and users are subject to reputation-based recognition within their respective communities.}

\textcolor{black}{In the field of software engineering, there are several traditional public code review platforms such as GitHub and Reddit. Stack Exchange is a Q\&A-based community that encourages quick feedback. Any developer can submit code snippets without the need to provide a complete project or engage in complex version control, making it highly accessible and ideal for extracting smaller, isolated code review requests. This contrasts with GitHub, which is more focused on project-driven collaboration, and Reddit, where the review process is often informal and may lack in-depth technical feedback. Unlike Reddit, where feedback is typically provided through informal post comments and may lack the necessary technical depth, Stack Exchange provides a more structured environment for feedback. Users are encouraged to give specific, detailed responses, and reviews often involve technical explanations and recommendations. This level of technical detail makes Stack Exchange an ideal environment for extracting meaningful data for public code review tasks.}

As we are dedicated to providing intelligent service methods for the code review process, the most popular public code review website on Stack Exchange, \textbf{Code Review}\footnote{https://codereview.stackexchange.com/}, is used as the data source for this study. The dataset from Code Review used for this study is the latest and largest original data version available from 2011 to 2023\footnote{https://archive.org/download/stackexchange/codereview.stackexchange.com.7z}. It records the main content and quality label of PCR requests in separate XML files.

Code Review website, which is positioned as peer code review conducted in the form of SQA (Software Quality Assurance). It aims to facilitate a quick and convenient code review process to accelerate the exchange of professional knowledge among peer developers. Any developer can initiate a review request within the community. Professional reviewers in the community then browse through and assess the necessity of the request. Subsequently, the developer selects appropriate tags for the request to match the corresponding reviewers. The dataset collects information on the requests content, the necessity of the request, and the corresponding tags.

\subsubsection{Data Preprocessing}

To align with the model input requirements of the baseline methods in Section~\ref{sec:5.3.2}, we have divided the overall data preprocessing into four steps as detailed below.

\textbf{(1) XML File Parsing:} There are two common parsing methods for XML files: 1. DOM parsing which involves converting an XML document into a DOM tree, allowing access to XML document information through a hierarchical object model, describing the information in a node tree format. This parsing method offers flexible access but may suffer from long processing times and memory overflow issues due to machine performance factors. 2. SAX parsing, unlike DOM parsing, operates in a sequential access mode for reading and writing XML files. When using this method to parse XML files, multiple event handling functions need to be written for various data in the file, but it consumes less memory. Due to hardware constraints, this study adopts SAX parsing, resulting in an HTML formatted file.

\textbf{(2) Data Cleaning and Code Snippet Extraction:} Since the XML file contains noise such as URLs, garbled characters, HTML tags, tabs, and line breaks, this step designs regular expressions for the HTML file obtained in Step (1) to remove noise and irrelevant characters. Specifically, the BeautifulSoup library is used to identify HTML tags and separate the text, code snippets, and tags in the request objects. For example, code snippets are collected using the HTML element <code>...</code>.

\textcolor{black}{\textbf{(3) Necessity Determination:} The distribution of request vote scores on the Code Review site is analyzed. It is observed that the distribution follows a power-law pattern, exhibiting a “long-tail” characteristic in which most posts receive relatively low scores, while a small proportion obtain significantly higher scores. In such distributions, the upper percentile (e.g., top 40–50\%) of items are often regarded as high-quality “core contributions” recognized by the community~\cite{newman2005power}. The analysis showed that, in the Code Review community, requests with a score of at least 4 corresponded precisely to the top 43\%. Therefore, requests with a vote score of at least 4 are classified as necessary requests, whereas those below this threshold are considered unnecessary requests. This resulted in a total of 26,632 necessary requests and 35,056 unnecessary requests.}

\textbf{(4) Tag Filtering and Cleaning:} The dataset contains a large number of low-frequency tags. The main reasons for the low frequency of tags are that they do not conform to developers' usage habits, have weak relevance to the topic, or have low discussion relevance. Previous studies have considered a tag as rare if its frequency is below a predefined threshold $\theta $~\cite{xu2021post2vec}. Intuitively, if a tag appears infrequently in such a large dataset, it is not widely recognized by developers. To mitigate the impact on the quality of training and testing data, these rare tags are removed. Consistent with previous research, this study sets the threshold $\theta $ for rare tags to 50, and all requests and tags that are rare or contain only rare tags are removed. The statistical information of the cleaned dataset is shown in Table~\ref{tab:2}. Finally, a dataset containing 76,161 requests and 424 tags is obtained.

\textbf{(5) Tag Processing and Vocabulary Addition:} Unlike other topic information stations, the tags in the Code Review involve specialized terms. In addition to common language tags such as ``java'' and ``python,'' there are also specific version-related tags such as ``c++13'' and ``python3.x''. The prompt learning method needs to learn and predict based on the words in the vocabulary. Since the model's tokenization automatically divides words based on symbols, tags are designed as expressions in natural language and added to the model's vocabulary by writing function rules. This preprocessing of tags is not necessary for baseline methods that do not require the use of a vocabulary for prediction.

Following previous research, we have utilized the common ten-fold cross-validation method to split the dataset. Specifically, we initially randomly have divided the dataset into 10 equally sized subsets. One of these subsets is selected as the test set. The remaining data samples are split into training and validation sets in an 8:1 ratio. This process is repeated during model training to test the model's effectiveness, and the results is averaged to assess model performance. In summary, the dataset is split into training, validation, and test sets in an 8:1:1 ratio. Detailed statistical information is provided in Table~\ref{tab:2}.

\begin{table*}
\small
  \caption{Public Code Review Dataset Statistics}
  \label{tab:2}
  \begin{tabular}{cccc}
    \toprule
     &Training Set &Validation Set &Test Set 
\\
    \midrule
    All Data& 61688& 6855 &7616
\\
    \midrule
 Quality Necessity Task & &  &
\\
 Necessary& 26632&2959 &3288
\\
 Unnecessary& 35056&3896 &4328
\\
 \midrule
 Tag Recommendation Task& & &
\\
    Average Number of Tags / Request Object& 2.95&2.89 &2.94
\\
    Average Number of Words / Tag& 2.89&2.89 &2.89\\
    \bottomrule
  \end{tabular}
\end{table*}

\subsubsection{Evaluation Metrics}

For the tag recommendation subtask, previous recommendation tasks in the SQA community~\cite{li2023dual} have used Precision@k, Recall@k and F1@k to evaluate performance. To maintain consistency, this study adopts the same evaluation metrics. Consistent with the review necessity subtask and the comment quality prediction subtask in the PCR domain~\cite{ochodek2022automated}, Accuracy and F1 are selected as metrics. The following will provide a detailed introduction to these evaluation metrics.

\textbf{(1) $Top-k$ Precision $Precision@k$:} This metric is commonly used in recommendation-related methods, measuring the proportion of true labels among the top $k$ labels recommended by a model. For a single request, the calculation of its $Top-k$ Accuracy is shown in Equation~\ref{eq:4} which computes $Precision@k_i$ by calculating the intersection of predicted labels $vk_i$ with the corresponding true labels $Ttag_i$ for each request object. Since a public code review dataset contains multiple requests, when calculating accuracy metrics for the entire dataset, the average accuracy results for all requests need to be taken into account. The specific calculation method is shown in Equation~\ref{eq:5} which represents the average accuracy of all request objects, calculated by first computing $Precision@k_i$ for each request object, and then averaging them to represent $Precision@k$.

\begin{equation}
\label{eq:4}
  Precision@k_i=\frac{\left| Ttag_i\cap vk_i \right|}{k}
\end{equation}

\begin{equation}
\label{eq:5}
  Presison@k=\frac{\sum\limits_{i=1}^{\left| R \right|}{Precision@k_i}}{\left| R \right|}
\end{equation}

\textbf{(2) $Top-k$ Recall $Recall@k$:} This metric is used to evaluate the model's ability to predict the complete set of true labels. Specifically, it measures the proportion of true labels that are contained in the predicted labels. The $Top-k$ Recall for a single request object, $Recall@k_i$, is calculated as shown in Equation~\ref{eq:6}. In this work, due to the true label length range being 3-5, and following the metrics settings of recommendation-related methods, $k$ is set to 3, 5, and 10 in this study. Similarly, since there are multiple request objects in the public code review dataset, the performance calculation of Recall metric $Recall@k$ for the entire dataset is shown in Equation~\ref{eq:7}. The average recall rate of all request objects is calculated to represent the model's performance on this metric.

\begin{equation}
\label{eq:6}
Recall@k_i=\left\{ \begin{array}{l}
	\left| \frac{Ttag_i\cap vk_i}{k} \right|,if\left| Tag_i \right|<k,\\
	\left| \frac{Ttag_i\cap vk_i}{Ttag_i} \right|,otherwise.\\
\end{array} \right. 
\end{equation}

\begin{equation}
\label{eq:7}
Recall@k=\frac{\sum\limits_{i=1}^{\left| R \right|}{Recall@K_i}}{\left| R \right|}
\end{equation}

\textbf{(3) $Top-k$ Macro-Average $F1@k$:} This metric is commonly used to evaluate a model's overall performance on precision and recall by calculating the weighted average of these two metrics. It provides a balanced view of a model's performance by considering both precision and recall. As described in Equation~\ref{eq:8}, the calculation method for the macro-average $F1@k_i$ of a single request object is similar to the previous two metrics, where $k$ is set to 3, 5, and 10 in this study. Similarly, for the overall macro-average metric $F1@k$ of the public code review dataset, its calculation is shown in Equation~\ref{eq:9}, derived from the average $F1$ value of all request objects.

\begin{equation}
\label{eq:8}
F1@k_i=2\times \frac{Precision@k_i\times Recall@k_i}{Precision@k_i+Recall@k_i}
\end{equation}

\begin{equation}
\label{eq:9}
F1@k=\frac{\sum\limits_{i=1}^{\left| R \right|}{F1@k_i}}{\left| R \right|}
\end{equation}

\subsection{Implementation Details}

\subsubsection{Implementation}

In practice, the KP-PCR framework can be applied to various pre-trained language models, such as BERT~\cite{9}, RoBERTa~\cite{hong2022commentfinder}, ALBERT~\cite{thongtanunam2022autotransform}, GPT~\cite{hong2022commentfinder}, etc. However, due to hardware limitations, this study uses the base version of BERT as the encoding model (BERT-base-uncased), and GraphCodeBERT as the code snippet encoder. In the experiments, the learning rate of the KP-PCR is set to 1e-5, and the number of epochs is set to 6.

This study implements KP-PCR based on the PyTorch framework,  and all experimental parameters are listed in Table~\ref{tab:3}. The maximum length of the text is set to 300, the length of the prefix vector is set to 50, and the code representation part of the prefix vector uses a code snippet with a length of 150 to generate the code snippet representation, which is then used for model training and learning with a length of 100. The code snippet text length that is not optimized for model training is 150.

\begin{table*}
\small
  \caption{Experimental Parameters}
  \label{tab:3}
  \begin{tabular}{cccc}
    \toprule
     Name& Parameter&Name&Parameter\\
     \midrule
    Max lenth& 300& Hidden size&768\\
 Prefix lenth& 50& Hidden layer& 12\\
 Code lenth& 150& Weight decay& 0\\
 Batch size& 4& Learning rate& 1e-5\\
 epoch& 6& Optimizer& Adamw\\
     \bottomrule
  \end{tabular}
\end{table*}

\subsubsection{Baselines}
\label{sec:5.3.2}

For overall comparisons, a series of state-of-the-art approaches are chosen as baselines for the tag recommendation subtask and the request necessityprediction subtask, respectively.

\textbf{(1) The request necessity prediction subtask}, CommentBERT~\cite{ochodek2022automated}, MetaTransformer~\cite{chen2022leveraging} and Commentfinder~\cite{hong2022commentfinder} are baseline methods, which have the same data input structure and task definition as review necessity prediction subtask.

\textbf{(2) The tag recommendation subtask}, TagDC~\cite{24}, PROFIT~\cite{27}, Post2Vec~\cite{xu2021post2vec}, PTM4Tag~\cite{29} and CDR4Tag~\cite{li2023dual} are compared for recommending tags to the SQA community.

Our previous study is also listed for comparison. In this paper, the source code provided by the baselines is used to fine-tune the parameters and obtain the optimal values to represent the experimental results of the baselines. For baselines which source code was not provided, this study reproduced the model design based on Pytorch framework. The related experiments in this study are conducted on four NVIDIA Titan XP GPUs. All hardware and software environments are shown in Table~\ref{tab:4}. 

\begin{table*}
\small
  \caption{Experimental Environment and Version Information}
  \label{tab:4}
  \begin{tabular}{cc}
    \toprule
       Name& Version Information\\
     \midrule
      GPU& NVIDIA Titan XP*4\\
   CPU&IntelR XeonR E5-2650 v4 @2.20GHz\\
   Memory Capacity&64G*4\\
   CUDA Version&11.0\\
 Python Version&3.5\\
 Pytorch Version&1.7\\
     \bottomrule
  \end{tabular}
\end{table*}

\section{Results}
\label{sec:6}

\subsection{Request Necessity Prediction Subtask (RQ1)}

Table~\ref{tab:5} shows the KP-PCR experimental results on the request necessity prediction compared with similarity-based and fine-tuning-based methods.

\begin{table*}
\small
  \caption{Experiment Results of Request Necessity Prediction Subtask}
  \label{tab:5}
  \begin{tabular}{cccc}
    \toprule
     Model& Accuracy&F1 Necessary& F1 Unnecessary\\
     \midrule
    CommentBERT~\cite{ochodek2022automated} & 0.608& 0.606 & 0.610\\
 MetaTransformer~\cite{chen2022leveraging}& 0.623& 0.687 & 0.525\\
 Commentfinder~\cite{hong2022commentfinder} & 0.616& 0.711& 0.426 \\
 UniPCR~\cite{chen2024unified}& \underline{0.798}& \underline{0.817}&\underline{0.775}\\
 \midrule
 \bfseries {Our KP-PCR} &  {\textbf{0.848}} & {\textbf{0.894}} & {\textbf{0.795}}\\
     \bottomrule
  \end{tabular}
\end{table*}

In Table~\ref{tab:5} it can be seen that KP-PCR achieves the best results compared to the other four baselines. Overall, the performance of the four methods on the non-necessary label F1 is lower than that on the necessary label F1. Based on actual data from public code reviews, this is because requests for unnecessary reviews appear in more diverse forms, such as asking questions about coding errors (which belong to coding issues rather than code review issues), submitting code that cannot be executed (which hinders reviewers from reading the code as a whole), and choosing inappropriate labels (which affects the professionalism of the review). In other words, requests that meet the review requirements have clearer expressions, executable code, and accurate identification of review issues.

\textbf{(1) Compared to baseline methods, KP-PCR shows improved performance in the review necessity prediction subtask.} As shown in Table~\ref{tab:5}, KP-PCR achieves a 2.3\%-8.4\% performance improvement over the second-best method (UniPCR) by introducing the code snippet optimization module. This demonstrates the effectiveness of our KP-PCR. By learning the representation of code snippets within the overall content through soft prompts in the prefix vectors, KP-PCR understands the differences in various code snippets under the review necessity criteria. During prediction, the representation in the learned code snippet prefix vectors enhances the prediction performance for review necessity, resulting in KP-PCR's improved performance in this subtask.

\textbf{(2) Optimizing code snippet representation is effective for the review necessity prediction task.} The experimental results indicate that a large amount of representation information contained in code snippets are not fully utilized in previous traditional architectures. CommentBERT~\cite{ochodek2022automated} and MetaTransformer~\cite{chen2022leveraging} use pre-trained language models to learn the entire request content but do not individually process or specially learn the code snippets. Commentfinder~\cite{hong2022commentfinder} treats the request body as a whole and uses similarity methods for matching. As mentioned earlier, non-essential tags in the review necessity subtask have more diverse forms and features in the data compared to essential tags, and these features are more prevalent in code snippets. The experimental results shown in Table~\ref{tab:5} show that extracting potential features from code snippets indeed helps accomplish the review necessity prediction task.

\textbf{(3) Introducing knowledge-guided prefix vectors in prompt learning is more conducive to learning code snippet representation.} KP-PCR achieves an understanding of the differences in various code snippets by learning the representation of code snippets within the overall content through prefix vectors in prompt learning. This means that the small prefix vectors, during the model training process, learn the initialization parameters, allowing the relevant information for determining necessity to be extracted from the complex information contained in the overall code snippets. During prediction, the representation in the learned code snippet prefix vectors enhances the prediction performance for review necessity. Compared to the baselines, using prompt learning prefix vectors to optimize code snippets is effective.

\subsection{Tag Recommendation Subtask (RQ1)}

This subsection compares our KP-PCR with state-of-the-art baselines on the tag recommendation subtask and analyzes the experimental results, in terms of Precision@k, Recall@k and F1@k. 

Tag recommendation subtask facilitates users to intuitively and quickly select the correct tags. As introduced in Section~\ref{sec:3.2}, users on code review websites can select 3-5 tags. Therefore, in the evaluation metrics, this study introduces @3, @5, and @10 as evaluation scopes. In addition, based on the analysis of the time complexity of our KP-PCR in Section~\ref{sec:4.3.1}, compared to the preliminary version of our study~\cite{chen2024unified}, our KP-PCR can improve the task accuracy performance without affecting the overall efficiency.

\begin{table*}
\small
  \caption{Experiment Results of Tag Recommendation Subtask}
  \label{tab:6}
  \begin{tabular}{c|ccc|ccc|ccc}
    \toprule
     \multirow{2}*{Model}  & \multicolumn{3}{c}{Precision} & \multicolumn{3}{c}{Recall} & \multicolumn{3}{c}{F1}\\
 & @3& @5&@10 & @3& @5&@10  & @3& @5&@10 \\
     \midrule
    TagDC~\cite{24}& 0.429& 0.311&0.189& 0.483& 0.570&0.678& 0.434& 0.387&0.288 \\
 PORFIT~\cite{27}& 0.511& 0.368&0.217& 0.571& 0.668&0.774 & 0.516& 0.456&0.330 \\
 Post2Vec~\cite{xu2021post2vec}& 0.550& 0.384&0.223 & 0.624& 0.707&0.799 & 0.558& 0.478&0.339 \\
 PTM4Tag~\cite{29}& 0.488& 0.339&0.199 & 0.557& 0.628&0.720 & 0.496& 0.423&0.303 \\
 CDR4Tag~\cite{li2023dual}& 0.564& 0.396& 0.227& 0.635& 0.724& 0.806& 0.572& 0.491&0.345\\
UniPCR~\cite{chen2024unified}& \underline{0.610}& \underline{0.436}&\underline{0.245}& \underline{0.688}& \underline{0.792}&\underline{0.870}& \underline{0.619}& \underline{0.541}&\underline{0.372}\\
 \midrule
 {{\bfseries Our KP-PCR}} & {\textbf{0.626}}& {\textbf{0.447}}&{\textbf{0.262}}& {\textbf{0.701}}& {\textbf{0.803}}&{\textbf{0.889}}& {\textbf{0.647}}& {\textbf{0.561}}&{\textbf{0.393}} 
\\
     \bottomrule
  \end{tabular}
\end{table*}

\textbf{(1) Compared to baseline methods, KP-PCR enhances the tag recommendation subtask.} Our KP-PCR outperforms baselines by 1.4\%-6.9\% in the tag recommendation task. In terms of handling code snippets, TagDC~\cite{24} and PORFIT~\cite{27} methods discard code snippets as noise, while Post2Vec~\cite{xu2021post2vec} and PTM4Tag~\cite{29} methods encode code snippets separately using the same approach as text segments. The unified modeling of UniPCR adjusts the model architecture through a prompt learning paradigm. Under the similar architecture, KP-PCR achieves better performance by designing a unique training method for code snippets while maintaining the advantage of the unified modeling method, which allows for modeling multiple subtasks.

\textbf{(2) Code snippet representation is also effective for the tag recommendation subtask.} Unlike other baseline methods, CDR4Tag~\cite{li2023dual} designs a matching module for code snippets to capture potential information, resulting in superior performance compared to other methods. This matching module is based on similarity design, completing the task by calculating the similarity between the code snippet and the request body. However, using this similarity-based approach is still insufficient for learning code snippet representation, as the matching module can only capture parts of the code snippet that match the text content, lacking direct learning of the overall code snippet content. The experimental results indicate that focusing on learning code snippet features rather than discarding code snippet content is necessary for the tag recommendation subtask. Compared to the proposed KP-PCR, it is evident that using a training method more aligned with the structure of code snippets leads to better performance in this task.

\textbf{(3) Introducing knowledge guidance is helpful for code snippet representation learning.} Compared to UniPCR~\cite{chen2024unified}, it is evident that knowledge-guided prompt learning is more effective in the tag recommendation subtask. As previously mentioned, CDR4Tag~\cite{li2023dual} constructs code snippets based on a matching module, but its shortcomings lie in the complexity and difficult migration of the module, which obscures the information within the code snippets themselves. UniPCR achieves competitive experimental results by narrowing the gap through task reconstruction, while KP-PCR achieves superior performance by introducing prefix vectors and program dependence graph representations of code snippets to learn the professional knowledge within the request content.

Overall, KP-PCR shows improvements across all metrics. Compared to the runner-up method UniPCR, it's evident that knowledge-guided prompting learning is more effective in the tag recommendation subtask. Moreover, CDR4Tag constructs code snippets based on a matching module, but its drawback lies in the complexity of the module, making it difficult to transfer and obscuring the information inherent in the code snippets themselves. While KP-PCR achieves even better performance by introducing prefix vectors and program dependence graph representations of code snippets for learning the professional knowledge embedded in the request content.

\subsection{Effects of Knowledge-Guided Propmt Template Designs (RQ2)}

\textcolor{black}{This section conducts experiments on the template designs for prompt learning, aiming to investigate the impact of using different template designs on the model's performance. Specifically, we detail the template designs for prompt learning as shown in Table~\ref{tab:7}.}

\textcolor{black}{\textbf{(1) Knowledge-Guided Soft Prompt.} We explore the impact of different types of knowledge guidance on model performance. Specifically, we analyze the effects of using (1) fine-tuned LLMs (DeepSeek, CodeLlama, LLaMA) and (2) general encyclopedic knowledge (Wikipedia) as knowledge guidance in two subtasks (request necessity prediction and tag recommendation). We fine-tune CodeLlama and LLaMA in the same manner as described in Section~\ref{sec:4.3.1}, using the same dataset and optimizing the fine-tuning parameters to their best values.}

\textcolor{black}{\textbf{(2) Code Prefix Prompt Construction.} We design experiments on different code parsing methods, converting code into DFG or PDG, to analyze their impact on the experimental results.}

\begin{table*}
\small
\captionsetup{labelfont={color=black}}
  \caption{\textcolor{black}{Template Designs for Prompt Learning}}
  \label{tab:7}
  \begin{tabular}{ccc}
    \toprule
     Name& Knowledge-Guided &Code Prefix\\
     \midrule
    Our KP-PCR & Fine-tuned DeepSeek-7B & PDG \\
    (a) & Fine-tuned CodeLlama-7B & PDG \\
    (b) & Fine-tuned Llama3-8B & PDG \\
    {\color{black}(c)} & {\color{black}Fine-tuned DeepSeek-7B} & {\color{black}DFG} \\
    (d) & Wikipedia & PDG \\
    (e) & Wikipedia & DFG \\
 (f) & $\times$ & DFG\\
 \bottomrule
  \end{tabular}
\end{table*}

\begin{table*}[h]
\small
\captionsetup{labelfont={color=black}}
  \caption{\textcolor{black}{Ablation Experiments of Request Necessity Prediction}}
  \label{tab:8}
  \begin{tabular}{cccc}
    \toprule
     Model& Accuracy&F1 Necessary&F1 Unnecessary\\
     \midrule
    Our KP-PCR& \textbf{0.848}& \textbf{0.894}&\textbf{0.795}\\
    (a)& 0.845& 0.892&0.793\\
    (b)& 0.836& 0.889&0.790\\
    {\color{black}(c)}& {\color{black}0.838}& {\color{black}0.890}&{\color{black}0.791}\\
    (d)& 0.835& 0.889&0.791\\
    (e) & 0.830& 0.885&0.785\\
    (f) & 0.745& 0.771& 0.712\\
 \bottomrule
  \end{tabular}
\end{table*}

\begin{table*}
\small
\captionsetup{labelfont={color=black}}
  \caption{\textcolor{black}{Ablation Experiments of Tag Recommendation}}
  \label{tab:9}
  \begin{tabular}{cccc}
    \toprule
     \multirow{2}*{Model}  & \multicolumn{3}{c}{Precision}   \\
 & @3& @5&@10 \\ 
     \midrule
    Our KP-PCR & \textbf{0.626}& \textbf{0.447}&\textbf{0.262}\\
  (a) & 0.625& 0.446&0.261\\
  (b) & 0.619& 0.441&0.253\\
  {\color{black}(c)} & {\color{black}0.621}& {\color{black}0.442}&{\color{black}0.255}\\
  (d) & 0.617& 0.442&0.251\\
  (e) & 0.613& 0.438&0.247\\
  (f) & 0.610& 0.436&0.245\\
\bottomrule
  \end{tabular}
  \begin{tabular}{cccc}
    \toprule
     \multirow{2}*{Model}   & \multicolumn{3}{c}{Recall}  \\
 & @3& @5&@10  \\
     \midrule
    Our KP-PCR & \textbf{0.701}& \textbf{0.803}&\textbf{0.889}\\
  (a) &  0.699& 0.800&0.887\\
  (b) &  0.698& 0.798&0.882\\
  {\color{black}(c)} &  {\color{black}0.699}& {\color{black}0.799}&{\color{black}0.884}\\
  (d) &  0.696& 0.796&0.875\\
  (e) &  0.692& 0.793&0.871\\
  (f) &  0.688& 0.792&0.870\\
\bottomrule
  \end{tabular}
\begin{tabular}{cccc}
    \toprule
     \multirow{2}*{Model}    & \multicolumn{3}{c}{F1} \\
 & @3& @5&@10  \\
     \midrule
    Our KP-PCR & \textbf{0.647}& \textbf{0.561}&\textbf{0.393} \\
  (a) & 0.646& 0.559&0.392 \\
  (b) & 0.642& 0.551&0.385 \\
  {\color{black}(c)} & {\color{black}0.644}& {\color{black}0.555}&{\color{black}0.388} \\
  (d) & 0.635& 0.548&0.379 \\
  (e) & 0.622& 0.543&0.373 \\
  (f) & 0.619& 0.541&0.372 \\
\bottomrule
  \end{tabular}
\end{table*}

\textcolor{black}{Table~\ref{tab:8} shows the experimental results on the necessity prediction subtask. Table~\ref{tab:9} shows the results of the ablation experiments on the tag recommendation subtask. }

\textcolor{black}{\textbf{(1)} Compared to template (a), (b), and (c), using the fine-tuned DeepSeek as an external knowledge source outperforms the fine-tuned CodeLlama, Llama3, or directly using Wikipedia. The fine-tuned DeepSeek improves performance by adjusting model parameters to the needs of specific tasks and providing more precise knowledge related to code review. }

\textcolor{black}{\textbf{(2)} The results of template (c) compared to our KP-PCR and the results of template (d) compared to template(e) show that if the code parsing method is changed from DFG to PDG, the accuracy can be improved by 1.5\%. PDG and DFG are two commonly used intermediate representation methods, modeling the structural information of programs from different perspectives. PDG models a program as a graph structure composed of statement nodes, with edges representing the composite relationship of control dependence and data dependence. DFG focuses more on depicting the data transfer paths among variables in the program, where nodes typically represent variables or operation units, and edges represent the flow of data. In the PCR tasks we focus on, the model needs to understand the execution logic of code snippets, judge the context where potential defects may occur, and make reasonable inferences about complex dependency relationships. We find that the representation method based on PDG performs more accurately on PCR tasks. This is mainly attributed to the fact that PDG can simultaneously capture the control structure and data transfer of a program, possessing higher structural integrity and semantic expressiveness, and providing more precise context information. }

\textcolor{black}{\textbf{(3)} The results of template (f) show that the accuracy reduces when the external knowledge is removed, indicating that external knowledge plays an important auxiliary role in this task.}

\subsection{LLM Baselines (RQ3)}

Recent advancements in large language models (LLMs) have demonstrated remarkable capabilities in code-related tasks. To rigorously evaluate the effectiveness of our proposed knowledge-guided prompt learning framework, we conducted a comparative study by fine-tuning Llama3-8B, CodeLlama-7B, and DeepSeek-7B for the two subtasks separately on our PCR dataset. These models were selected for their proven adaptability to code comprehension tasks.

The fine-tuning process utilized a training set of 61,688 samples and a validation set of 6,855 samples, with task labels aligned to necessity prediction and tag recommendation subtasks. We employed a learning rate of $
2\times 10^{-5}$, a batch size of 8, and trained for 10 epochs using mixed-precision training. Both models were optimized with cross-entropy loss and early stopping based on validation performance.

The performance of the fine-tuned models was benchmarked against our proposed KP-PCR framework. Unlike zero-shot or prompt-based evaluations, this approach allows for a fair comparison by leveraging domain adaptation through fine-tuning. As shown in Table~\ref{tab:rn_llm}, our experiments reveal that in the request necessity prediction subtask, LLMs are behind KP-PCR by 22.4\%-11.9\%. As shown in Table~\ref{tab:tr_llm}, LLMs lag behind KP-PCR by 34.2\% and 12.6\% in the tag recommendation subtask. These disparities demonstrate that, while LLMs exhibit baseline competence, their reliance on pure textual patterns fails to address the semantic complexity inherent in PCR tasks. 

\begin{table*}[h]
\small
  \caption{Experiment Results of Request Necessity Prediction Subtask}
  \label{tab:rn_llm}
  \begin{tabular}{cccc}
    \toprule
     Fine-Tuned LLM& Accuracy&F1 Necessary& F1 Unnecessary\\
     \midrule
    Llama3-8B & 0.654& 0.583& 0.698\\
 CodeLlama-7B& 0.692& 0.617& 0.709\\
 DeepSeek-7B& 0.742& 0.763& 0.721\\
 {\bfseries Our KP-PCR} & \textbf{0.848} & \textbf{0.894} & \textbf{0.795}\\
     \bottomrule
  \end{tabular}
\end{table*}

\begin{table*}[h]
\small
  \caption{Experiment Results of Tag Recommendation Subtask}
  \label{tab:tr_llm}
  \begin{tabular}{c|ccc|ccc|ccc}
    \toprule
     \multirow{2}*{Fine-Tuned LLM}  & \multicolumn{3}{c}{Precision} & \multicolumn{3}{c}{Recall} & \multicolumn{3}{c}{F1}\\
 & @3& @5&@10 & @3& @5&@10  & @3& @5&@10 \\
     \midrule
    Llama3-8B & 0.411& 0.334&0.215& 0.321& 0.452&0.575& 0.346& 0.312&0.259 \\
 CodeLlama-7B & 0.546& 0.391&0.207& 0.487& 0.601&0.762 & 0.518& 0.463&0.352 \\
DeepSeek-7B & 0.583& 0.457&0.281& 0.535& 0.634&0.781 & 0.529& 0.508&0.446 \\
 {\bfseries Our KP-PCR} & \textbf{0.626}& \textbf{0.447}&\textbf{0.262}
 & \textbf{0.701}& \textbf{0.803}&\textbf{0.889}
 & \textbf{0.647}& \textbf{0.561}&\textbf{0.393} 
\\
     \bottomrule
  \end{tabular}
\end{table*}

\section{Related Work}
\label{sec:7}

This section introduces and analyzes related work from five perspectives: tag recommendation task on software Q\&A sites, public code review, prompt learning in software engineering, code language model, and comparison with our preliminary study. 

\subsection{Tag Recommendation Task on Software Q\&A Sites}

Since code reviews are conducted in the form of software Q\&A sites, developers need to select appropriate tags to label their review requests during the submission stage. These tags are also used to recommend requests to suitable reviewers. Therefore, the accuracy of these tags is directly related to the final review quality.

Tag recommendation techniques have become an important research focus for software Q\&A sites aiming to provide better services for code review. Numerous studies have been proposed for software information sites~\cite{baum2016need}. In the early stages, some research works assumed that code snippets were noise and removing them during preprocessing would improve the model's performance such as TagDC~\cite{24} and PROFIT~\cite{27}.

Unlike previous works, the Post2Vec model~\cite{xu2021post2vec} preserves code snippets and learns the content of posts, including titles, descriptions, and code snippets, seamlessly integrating them to learn a unified representation. The authors of the PTM4Tag model~\cite{29} argue that poorly chosen tags often introduce additional noise and redundancy, leading to problems such as tag synonyms and tag explosion. Domain knowledge is not well studied in previous studies which may reduce the quality of tag recommendation.

\subsection{Code Review}

Due to the extensive review work involved, conducting code reviews has become one of the biggest challenges facing the industry~\cite{balachandran2013reducing}. Specifically, code reviews are labor-intensive because reviewers must manually inspect the code and provide written feedback. To alleviate the cost of review requests, considerable attention and research has been devoted to various automation techniques. 

For instance, Thongtanunam et al.~\cite{thongtanunam2022autotransform} proposed a code transformation service that automatically converts source code into improved versions that have been reviewed and approved. 
Kudrjavets et al.~\cite{kudrjavets2022mining} investigated opportunities for potential speedups in code by analyzing the switch from manual to automatic merging, which they found could enhance code velocity. 

Tufano et al.~\cite{DBLP:conf/icse/Tufano23} used pre-trained transformers like T5 to automate specific code review tasks such as revising code and generating review comments, aiming to improve code quality and reduce the time developers spend on manual reviews. 
Hijazi et al.~\cite{DBLP:journals/tse/HijaziDCCBMCCM23} proposed an approach to evaluate the quality of modern code reviews using biometric data, such as heart rate variability and pupillary response, to assess reviewers' cognitive load and predict the effectiveness of bug detection, providing feedback on code regions that may require further review. This approach requires gathering the reviewer's physiological signals such as Heart Rate Variability (HRV) and pupillary responses, which are difficult to retrieve in public code reviews. 
Hajari et al.~\cite{DBLP:journals/tse/HajariMMR24} proposed a novel code review recommendation system, SofiaWL, which balances expertise, workload distribution, and turnover risk to improve code review efficiency and mitigate knowledge loss in software projects.

These methods are primarily designed around the comments and code changes of individual reviewers, which may limit their effectiveness in the context of community-driven public code reviews.

\subsection{Pubilc Code Review}

Unlike internal code reviews within teams, public code reviews are implemented in the form of software question-and-answer sites. Such as StackExchange Code Review, a public, community-driven platform, where developers seek feedback from a broad audience. This form, characterized by a large user base and rapid response, also has features such as high knowledge dissemination. However, it also introduces factors that can affect quality and efficiency, primarily related to developers. Public code reviews on websites often involve more complex and dynamic interactions, including discussions, negotiations, and multiple rounds of revisions. 

Since the quality and responsiveness of a public code review are typically determined by developers, this process is often carried out manually, making it difficult to achieve convenience and quality assurance. Therefore, with the popularity of this review format, the demand for a shift from manual to intelligent services in public code reviews has also arisen. The process of ensuring review quality in public code reviews can be seen as a review necessity prediction and a tag recommendation task.

Li et al.~\cite{li2022automating} proposed that by systematically considering the tasks involved in the entire review process and modeling them, the time and resource costs of the review process can be greatly reduced. Existing research works mostly employ fine-tuning paradigm modifications of model architectures to ensure the model's performance in terms of accuracy and other relevant metrics for the tasks. This paradigmatic method can accomplish individual tasks with individual models. Table~\ref{tab:13} shows the statistical information on model construction for the aforementioned two subtasks. It can be observed that although different modeling methods are used for different tasks, most existing research is based on pre-trained language models.

\begin{table*}
\small
  \caption{The Model Classifications Used in Existing Methods}
  \label{tab:13}
  \begin{tabular}{ccccc}
    \toprule
     Name & Pre-trained Language & CNN & DAG & Similarity\\
     \midrule
    TagDC~\cite{24} &  & $\surd$& &\\
    PORFIT~\cite{27} &  & & $\surd$&\\ 
    CommentBERT~\cite{ochodek2022automated} & $\surd$ & & &\\
    MetaTransformer~\cite{chen2022leveraging} & $\surd$ & & &\\
    Commentfinder~\cite{hong2022commentfinder}&  & & & $\surd$ \\
    Post2Vec~\cite{xu2021post2vec}& $\surd$ & & &\\
    PTM4Tag~\cite{29}&  & $\surd$ & &\\
    CDR4Tag~\cite{li2023dual}& $\surd$ & & &\\
    UniPCR~\cite{chen2024unified}& $\surd$ & & &\\
 \bottomrule
  \end{tabular}
\end{table*}

\subsection{Prompt Learning in Software Engineering}

Fine-tuning paradigm, where models tailored to specific tasks are trained only on target task's input-output example datasets, has long played a central role in many machine learning tasks~\cite{lu2023llama,59,60}. Currently, prompt learning can be divided into two types: hard prompts and soft prompts.

Currently, both have found rich applications in various natural language processing tasks. For example, Han et al.~\cite{59} augmented MLM with rule-based designs to decompose challenging tasks into several simpler sub-tasks. Similarly, Li et al.~\cite{60} introduced prefix tuning, which optimizes small vectors of task-specific parameters, achieving similar performance to fine-tuning by learning only 0.1\% of the parameters.

In the field of software engineering, prompt tuning has seen increasing applications. Due to the lightweight and efficient nature pursued in software engineering tools, existing research efforts have gradually attempted to use methods similar to prompt tuning as alternatives to fine-tuning. For example, Lu et al.~\cite{lu2023llama} proposed an LLaMA-Reverwer framework, which leverages mainstream large language models and adopts a parameter-efficient fine-tuning method to address the computational challenges of fine-tuning large language models. Similarly, the latest tool, POME, introduced by Huang et al.~\cite{62}, infers non-fully qualified type name syntax and usage knowledge encapsulated in prompt-tuned code-masked language models through a fill-in-the-blank strategy.

\subsection{Code Language Model}

To better capture the inherent structure of code, GraphCodeBERT proposed by Guo et al.~\cite{63} and TRACED proposed by Ding et al.~\cite{ding2024traced} can be used. 
GraphCodeBERT demonstrated significant improvements in adapting to downstream tasks compared to other pre-trained language models across multiple code-related tasks. GraphCodeBERT, as such a programming language-based pre-trained model, is constructed inspired by natural language pre-trained models, enhancing model training by incorporating code-specific pre-training tasks on top of natural language pre-training. Existing pre-trained models treat code snippets as token sequences, ignoring the inherent structure in code that provides crucial semantics for enhanced code understanding. TRACED, an execution-aware pre-training strategy for source code, to teach code models the complicated execution logic during the pre-training, enabling the model to statically estimate the dynamic code properties without repeatedly executing code during task-specific fine-tuning. GraphCodeBERT utilizes data flows in the pre-training phase, representing the semantic-level structure of code, encoding relationships between variable values. This semantic-level structure is less complex and does not introduce unnecessary deep hierarchical structures, making the model more effective. GraphCodeBERT is used as pre-trained model in our study. 

\subsection{Comparison with Our Preliminary Study}
\label{sec:7.5}

In our preliminary study~\cite{chen2024unified}, a unified model is proposed for the request necessity prediction subtask and the tag recommendation subtask. The study in this paper represents a significant expansion of that study in several respects. Firstly, we introduce a knowledge guidance base by soft prompt to generate prefix vectors which guide the model to understand the meaning of the code snippet. Secondly, we design the knowledge-guided prompt learning on request quality assurance task to show that the knowledge guide can further improve the request quality of PCR. Thirdly, we set up a prompt template design study to explore the influence of knowledge guidance. Further, we analyze the time complexity to prove that our KP-PCR can improve task accuracy while maintaining overall efficiency. Finally, we design a case study to investigate the performance of our prompt templates on large-scale language models. In accordance with these additional results, the entire paper has been thoroughly updated to reflect the new holistic study.

\section{Conclusions}
\label{sec:8}

Through investigating and summarizing the relevant tasks of team code review and software question-and-answer sites, and combining them with the process of public code review, this study explores intelligent service methods for the necessity prediction task and tag recommendation task of public code review from the perspective of developers. Addressing the issue of semantic understanding of professional knowledge in the request content, we introduce knowledge guidance and adopt a processing and training method that conforms to the semantic structure of code snippets to achieve a more comprehensive learning of professional knowledge. We propose KP-PCR to optimize the learning of professional knowledge in public code review. In public code review, request content includes both text and code snippets. The text contains vocabulary related to professional knowledge, while the unique structure of code snippets contains representations of content strongly related to review quality as the main part of the content. By processing code snippets into program dependence graphs and guiding the model to learn professional knowledge with knowledge-guided prefix vectors, KP-PCR achieves significant improvements in both subtasks. In the experimental results of the two subtasks, KP-PCR shows an improvement of 5.3\% to 27.9\% in the necessity prediction subtask and an improvement of 2.4\% to 31.4\% in the tag recommendation subtask.

\section*{Data Availability} The replication package of this study has been made available at \url{https://github.com/WUT-IDEA/KP-PCR}.

\begin{acks}
This work has been partially supported by the National Natural Science Foundation of China (NSFC) with Grant Nos. 62276196 and 62172311. The numerical calculations in this paper have been done on the supercomputing system in the Supercomputing Center of Wuhan University.
\end{acks}

\bibliographystyle{ACM-Reference-Format}
\bibliography{references}

\end{document}